\documentclass{article}
\usepackage[T1]{fontenc} % add special characters (e.g., umlaute)
\usepackage[utf8]{inputenc} % set utf-8 as default input encoding
\usepackage{ismir,amsmath,cite,url}
\usepackage{graphicx}
\usepackage{color}
\def\lbd{}	% Flag to use correct LBD settings in the paper, please do not modify this line

\usepackage{lineno}
\usepackage{url}

\title{JAMMIN-GPT: Text-based Improvisation using LLMs in Ableton Live}

\multauthor
{Sven Hollowell \hspace{1cm} Tashi Namgyal \hspace{1cm} Paul Marshall} {Department of Computer Science, University of Bristol, Bristol, UK\\
{\tt\small sven.hollowell@bristol.ac.uk}
}

\def\authorname{S. Hollowell, T. Namgyal, and P. Marshall}

\begin{document}

\maketitle
\begin{abstract}

We introduce a system that allows users of Ableton Live to create MIDI-clips by naming them with musical descriptions. Users can compose by typing the desired musical content directly in Ableton's clip view, which is then inserted by our integrated system. This allows users to stay in the flow of their creative process while quickly generating musical ideas. The system works by prompting ChatGPT to reply using one of several text-based musical formats, such as ABC notation, chord symbols, or drum tablature.
This is an important step in integrating generative AI tools into pre-existing musical workflows, and could be valuable for content makers who prefer to express their creative vision through descriptive language. Code is available at\footnote{\url{https://github.com/supersational/JAMMIN-GPT}}.
\end{abstract}

\section{Introduction}\label{sec:introduction}

Digital Audio Workstations (DAWs) are the primary tool for music production. Ableton Live is a popular DAW that uses a clip-based workflow, where users create and edit musical ideas in short clips that can be launched in a grid (rows are "scenes", columns are "tracks"). Typically, MIDI-clips would be created by drawing notes in a piano roll editor, or by playing notes on a MIDI keyboard. 

Music creation can be done in a variety of ways but is widely seen to be an iterative process \cite{huang2020song, garcia2014structured}. For example, with a "flare and focus" approach \cite{buxton2010sketching} where musicians successively expand upon and refine ideas. This often includes communicating with other musicians who have different roles, interaction styles and levels of grounding \cite{bryankinns2021exploring}.According to Sawyer \cite{sawyerCreativity} ``In group creativity, the performance must be constantly negotiated and constructed from moment to moment''. However traditional DAWs are tailored for solo creators, lacking the inherent spark of creativity that collaboration with a creative partner allows for. Using a virtual collaborator moves the user into the role of a producer who might give more general directions, e.g. ``play that again but with more energy''. This process is often done in natural language, and so LLMs are a natural go-to for emulating this part of the co-creation process. 

Another important aspect of composition is 'flow' \cite{mihaly2008flow}, where musicians enter a state of continuous inspiration. However, it is easier to break this state than to enter it and so AI tools that take users away from creative to administrative tasks should be avoided. For example, having to load special environments or wait for long training and/or inferences times. One solution is to embed interfaces within existing workflows, such as DAWs. We therefore propose JAMMIN-GPT, a natural language interface for music generation embedded within a DAW.

\section{Related Work}\label{sec:related_work}

There are many examples of AI-based composition tools being imported into DAWs as plugins, such as DrumNet \cite{lattner2019high}, MMM4Live \cite{ens2020mmm} and Magenta Studio \cite{magenta-studio}. These allow creation of MIDI clips by using a generative model such as MusicVAE \cite{MusicVAE}, which can generate novel MIDI clips or variations of existing clips. The central disadvantage of this approach is that the user cannot specify the musical content of the clip, since the clip is generated from a latent space that is not always interpretable to humans. Another disadvantage is that the interface makes it awkward to select input and output clips, which must be selected from dropdown menus. 
Our approach improves on this as it built directly into the clip-view of Ableton Live.

Text-conditioned generative models for audio have recently enabled users to generate clips of music from a text description, such as AudioLM \cite{borsos2023audiolm}, AudioLDM \cite{liu2023audioldm} and MusicGen \cite{copet2023simple}. However, generating directly in the audio domain makes it difficult for users to tweak model outputs compared to symbolic approaches.

Language models have been used to generate symbolic music in various ways. Models can be trained solely on music data, solely on natural language, or trained on natural language and then fine-tuned on music data. For example Music Transformer \cite{huang2019music} uses an LLM-like architecture trained specifically for music generation, but is smaller in scale than natural language models so is less expressive. 

There are many kinds of symbolic music representation, such as MIDI, that are not text-based and so are not present in the data used to train LLMs. However, these can be converted to a text format and used to fine-tune LLMs, for example fine-tuning GPT-2 to piano music \cite{banar2022systematic}. ChatGPT and GPT-3 models have been used by Tomoki \cite{Okuda2023Investigation} to generate code for the TidalCycles live coding environment, by fine-tuning it on example pairs of text-descriptions and code. 
GPT-3 has also been used by Zhang et. al. \cite{LLMsAreDrummers} to generate drum pattern continuations from a starting snippet. Zhang et. al. used a similar approach to Tomoki, but used a larger dataset of drum patterns. They demonstrate that the generated patterns are both musically plausible and diverse, being different from any of the training examples.

 % Paragraph on audio generation with text conditioning, such as MusicLM, MusicCraft, AudioGen, MusicLDM. Mention CLIP and CLAP, datasets such as MusicCaps. 

\section{Design}\label{sec:design}

At a high level, our main system operates by reacting to messages from Ableton Live via the OSC protocol. When a user creates and names an empty MIDI clip, we use the name of the clip as part of a prompt for ChatGPT (GPT-4-turbo \cite{OpenAIModels}). We get ChatGPT to create MIDI data by prompting it to respond in one of several text-based musical formats, such as ABC notation, chord symbols, or drum tablature. Our system processes ChatGPT's response, converting it back into MIDI, and inserting it into the clip.

\subsection{ChatGPT Abilities}\label{subsec:prompting} 
ChatGPT is able to produce music in a variety of text-based music formats, which were discovered by asking it for a list of formats it knows. This is an important choice for generating high quality output, since ChatGPT will have seen examples of different styles of music in different formats. For example, ABC music notation typically represents folk music, so the generated music will reflect this bias.

\subsection{Ableton to Python Interface}\label{subsec:ableton}
Ableton Live supports the use of MIDI remote scripts to control the DAW from an external device. We provide a remote script for the user to install, which provides an OSC interface from Ableton Live.

Using Python, we can control many features of Ableton Live, such as reading or launching clips, or changing the BPM. Our Python script polls Ableton Live for changes in clip names, and sends this information along with relevant musical context to the ChatGPT model. We use Ableton Live's clip colour feature to indicate that a clip is being generated.
When the generation is complete we parse the MIDI from ChatGPTs response and insert it into the clip, changing color to indicate it is completed. We also allow for editing existing clips. If a clip already containing MIDI is renamed, we prompt ChatGPT with the content of the original clip and ask it to alter the MIDI based on the prompt.
Instrument choice is left to the user. The system uses the name of the clip's track as part of its prompt, so it has information on which instrument is being used.

\begin{figure}
 \centerline{\framebox{
 \includegraphics[width=\columnwidth]{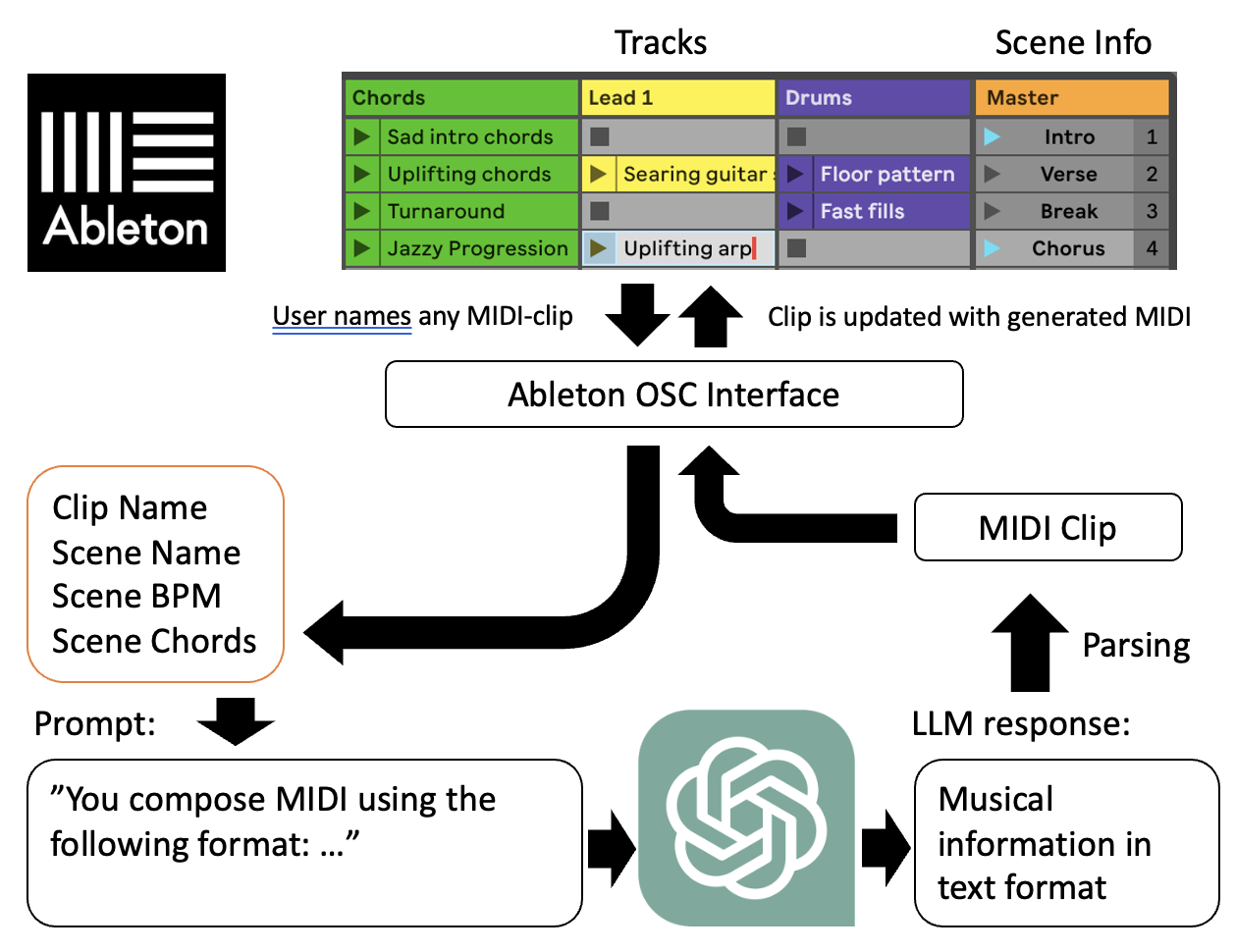}}}
 \caption{System overview, from user input to MIDI output. The user interaction consists of selecting clips and typing a description. 
 % JAMMIN-GPT completes the rest.
 JAMMIN-GPT automates the remaining workflow.
 }
 \label{fig:system}
\end{figure}

\subsection{MIDI Parsing}
We determine the output format based on the prompt. One mode uses simple keyword extraction i.e. it uses chord-symbols if the prompt contains the word "chord" or "chords". The other method is to prompt ChatGPT to choose the format by prompting it to first "choose" a format and then generate the output. This is not optimal since the model often favours chord-symbol notation, even when it's not suitable. For instance, when asked for a "funky bassline" it will choose chord-symbol notation which lacks the required syntax for rhythm.

\section{Discussion}
Our system has been tested by a small number of Ableton users. We observed that users were able to quickly begin using and prompting the system, since the interface is familiar to them. Some users prompted ChatGPT with audio based descriptions that are not representable with MIDI such as "reverb-y" or "grainy synth sound". These kinds of features can not currently be controlled by the model.

Musicality of the generated MIDI varies by format and style. Generally, ChatGPT generates meaningful and prompt-sensitive chord progressions which are of high quality. However, its lack of exposure to styles other than folk in the ABC format biases the model towards folkier styles regardless of the prompt. In future we hope to improve its ability to generate ABC notation by converting a dataset of various styles of music into ABC and fine-tuning the model on it. Since the backend is not fixed to use ChatGPT only, we could also fine-tune a model such as LLaMA \cite{touvron2023llama} or incorporate other models such as MusicVAE \cite{MusicVAE} to refine the output of ChatGPT. The LLM can also be swapped or upgraded as future models are released.

We note that existing methods for using AI within a DAW are limited in both their UI and the fact that natural language description cannot be used. To this end we designed a system that is both intuitive to use, and uses ChatGPT so that MIDI clips can be generated from natural language musical descriptions. We see the clip-based text-to-MIDI interface as a modality worth exploring further, since it allows for quick experimentation and iteration.

\section{Acknowledgments}

Sven Hollowell and Tashi Namgyal are supported by the UKRI Centre for Doctoral Training in Interactive Artificial Intelligence (EP/S022937/1).

\bibliography{report}

\end{document}